\documentclass[conference,compsoc]{IEEEtran}
\IEEEoverridecommandlockouts

\usepackage{cite}
\usepackage{amsmath,amssymb,amsfonts}
\usepackage{algorithmic}
\usepackage{graphicx}
\usepackage{textcomp}
\usepackage{xcolor}
\usepackage{url}
\usepackage{enumitem}

\newboolean{usetodonotes}
\setboolean{usetodonotes}{false}

\usepackage[textsize=scriptsize]{todonotes}
\ifthenelse{\boolean{usetodonotes}}{
  \setlength{\marginparwidth}{1cm}
  \paperwidth=\dimexpr \paperwidth + 6cm\relax
  \oddsidemargin=\dimexpr\oddsidemargin + 3cm\relax
  \evensidemargin=\dimexpr\evensidemargin + 3cm\relax
  \marginparwidth=\dimexpr \marginparwidth + 3cm\relax
  \presetkeys
  {todonotes}
  {noinline,caption={}}{}
}{
  \presetkeys
  {todonotes}
  {disable}{}
}

\def\BibTeX{{\rm B\kern-.05em{\sc i\kern-.025em b}\kern-.08em
    T\kern-.1667em\lower.7ex\hbox{E}\kern-.125emX}}

\let\oldparagraph\paragraph
\renewcommand{\paragraph}[1]{\oldparagraph{\textbf{#1}}}

\begin{document}

\title{Not Discrete Enough: On the Inherent Insecurity of dTPMs for Measured Boot}

\author{\IEEEauthorblockN{Christian Werling}
\IEEEauthorblockA{cwerling@sect.tu-berlin.de\\
SecT, TU Berlin\\
Berlin, Germany}
\and
\IEEEauthorblockN{Tahmid Zahin}
\IEEEauthorblockA{zahin@campus.tu-berlin.de\\
TU Berlin\\
Berlin, Germany}
\and
\IEEEauthorblockN{Jean-Pierre Seifert}
\IEEEauthorblockA{
jean-pierre.seifert@tu-berlin.de \\
SecT, TU Berlin \& Fraunhofer SIT\\
Berlin, Germany \\}
}

\maketitle

\makeatletter
\begingroup
\renewcommand\thefootnote{}
\let\@makefnmark\relax
\let\@IEEEcompsocmakefnmark\relax
\footnotetext{This is the authors' version of this work. It is posted here for personal use, not for redistribution. The definitive Version of Record was published in the \emph{2025 Annual Computer Security Applications Conference Workshops (ACSAC Workshops)}, pp.~546--550, DOI: \url{https://doi.org/10.1109/ACSACW69556.2025.00063}.

\copyright~2025 IEEE. Personal use of this material is permitted. Permission from IEEE must be obtained for all other uses, in any current or future media, including reprinting/republishing this material for advertising or promotional purposes, creating new collective works, for resale or redistribution to servers or lists, or reuse of any copyrighted component of this work in other works.}
\endgroup
\makeatother

\begin{abstract}
    Measured Boot, a mechanism enabled through Trusted Platform Modules (TPMs), is commonly used for passwordless protection of data-at-rest, aiming to protect data when the device is lost or stolen.
    Microsoft's standpoint is neutral on which way a TPM should be implemented:
    Firmware-based TPMs (fTPMs) are viewed as more economical but less secure.
    Despite the inherent susceptibility to bus sniffing attacks, discrete TPMs (dTPMs) are still seen as the gold standard, as many deliver better on-paper tamper resistance.
    It is often argued that attacks against the bus can be mitigated by bus encryption and, ideally, mutual authentication between the CPU and TPM.

    This position paper aims to emphasize another inherent, difficult-to-mitigate attack against dTPMs that was originally shown against a TPM 1.1 over 20 years ago:
    We demonstrate that even brief physical access to a TPM 2.0 and the ability to boot from an attacker-controlled system enable an attacker to reset and replay arbitrary measurements, thereby allowing an attacker to unseal, for example, a disk encryption key solely protected by the TPM.
    While there have been attacks against fTPMs, too, we argue that their practical attack surface is fundamentally smaller.
    Bus protection techniques can be used to protect dTPMs, but only guard against passive attacks.
    After all, we argue that, from a security standpoint, firmware TPMs, or any TPM internal to the SoC, are superior to discrete (external) ones.
    Lastly, in order for dTPM-based setups to provide meaningful protection of sealed secrets, configurations must require a user-provided PIN or password along with the Measured Boot configuration.
    \end{abstract}


\section{Introduction}
Full disk encryption (FDE) has become a critical security measure for protecting data at rest, with Microsoft BitLocker being the go-to solution for the large share of Windows-based systems.
As the cryptographic primitives supporting modern FDE solutions  (e.g., AES XTS) are regarded as secure in industry and academia, their security boils down to the secrecy of the disk key.
A TPM can alleviate the user's burden of memorizing and entering a secure password at pre-boot by sealing this key under specific policies.
Many deployments rely on these TPM-only configurations for a seamless boot into the operating system's login screen, where its password prompt protects the decrypted data from unauthorized (online) access eventually.

\subsection{Background}

\paragraph{Platform Configuration Registers (PCRs)} PCRs are 256-bit registers that maintain cryptographic summaries of platform state through hash-chain semantics. The \texttt{TPM2\_PCR\_Extend} operation concatenates new measurements with existing PCR values: $\text{PCR}^{\text{new}}_i = H(\text{PCR}^{\text{old}}_i \| m)$, where $H$ is the hash algorithm and $m$ is the new measurement. This creates an append-only chain that is assumed to reset only with a platform reset.

For BitLocker's TPM-only protector, two PCRs are critical: PCR 7 measures the UEFI firmware's Secure Boot configuration and state, while PCR 11 is used in its null state and is extended into \emph{after} Windows has unsealed the disk to prevent other software from unsealing the key data~\cite{itm4n_bitlocker_deepdive}.

\paragraph{Sealed Storage and Policy Sessions} TPM 2.0's sealed storage allows binding secrets to specific platform states using policy sessions. For BitLocker, the Volume Master Key (VMK) is sealed with a policy requiring the validation of PCRs 7 and 11. The unsealing process involves: (1) initiating a policy session, (2) satisfying PCR policy requirements, and (3) executing \texttt{TPM2\_Unseal} to release the VMK.

\paragraph{TPM Form Factors}
TPMs are implemented in two primary forms: \emph{Discrete TPMs} (dTPMs, sometimes called \emph{external TPMs}) are separate chips on the motherboard. In contrast, internal/integrated TPMs, such as \emph{firmware TPMs} (fTPMs), are implemented in software or hardware inside the System on a chip (SoC).
 While dTPMs offer stronger tamper resistance and are certified under Common Criteria, they require external communication buses that create potential attack vectors~\cite{tcg2015brief}.

\subsection{TPM Reset Attack}

Our demonstrated attack is not new, but was already published in a paper by Bernhard Kauer ~\cite{kauer2007oslo} where he claims: "In July 2004 we discovered that setting the reset bit in a control register of a v1.1 TPM resets the chip without resetting the whole platform."
He goes on: "As it results in default PCR values, this breaks the remote attestation and sealing features of those chips: Any PCR value can be reproduced [...]".

In this position paper, we reproduce this over 20-year-old attack on two recent systems with TPM 2.0, one laptop and one desktop system, both running Windows 11 with TPM-only protected drives (the default on Windows 11).
Given brief physical access to the motherboards and the ability to boot from a USB live Linux, it allows us to unseal a BitLocker Volume Master Key successfully and, with it, decrypt the whole Windows volume from within Linux.

Our demonstrated attack targets a specific application of \emph{Measured Boot}~\cite{tcg_pcclient_spec}, namely Full Disk Encryption (FDE) with Microsoft BitLocker. However, we have internally verified that the attack works against Linux-based FDE with LUKS, too. Furthermore, we argue that the same attack can be mounted for other secrets protected by Measured Boot based on dTPMs (e.g., VPN keys).

\section{Related Work}

\paragraph{Side-channel Attacks Against TPMs}
TPM-FAIL~\cite{moghimi_tpm-fail_2020} performs black-box timing analysis of TPM 2.0 devices to find secret-dependent execution times during signature generation. These timing leakages were discovered on both Intel fTPMs and discrete TPMs by STMicroelectronics. Both vendors have released firmware updates addressing these vulnerabilities.

\paragraph{Power Management Attacks}
In 2018, Han et al. reported two types of TPM attacks related to power management, finding ways to reset and forge TPM PCR values~\cite{han_bad_2018}. One vulnerability targets a design flaw in the TPM 2.0 specification, while the other exploits an implementation flaw in \emph{tboot}, the most popular measured launch environment used with Intel's TXT. While patches were provided for the latter, the former was reported to Intel, Dell, Gigabyte, and Asus.

\paragraph{Hardware Bus Sniffing Attacks}
Recent work demonstrated practical attacks against dTPM-protected BitLocker using low-cost hardware ($\sim$\$5 Raspberry Pi Pico) to intercept VMK transmission over unencrypted LPC/SPI buses~\cite{stacksmashing-tpm-attack}. 
This attack requires physical access but completely bypasses software protections by capturing the cleartext key during TPM-to-CPU transmission.

\paragraph{Firmware TPM Fault Injection}
While fTPMs are immune to bus sniffing, voltage fault injection attacks against AMD's fTPM have shown that the underlying secure processor can be compromised, rendering both TPM-only and TPM+PIN protectors ineffective~\cite{faultpm}.

\paragraph{Software-based TPM Reset Attacks}
Research by Hacky Solutions (2024) showed that dTPMs can be reset through software manipulation alone, thus enabling replay of legitimate boot sequences to forge required PCR states~\cite{hacky-solutions-tpm-attack-2024}. This work builds on Kauer's foundational research on TPM reset vulnerabilities~\cite{kauer2007oslo}, demonstrating that disorderly shutdown states can enable \texttt{TPM2\_Startup(CLEAR)} commands to reset PCRs.

\section{Experimental Results}

All code to reproduce the attack is accessible in a dedicated GitHub repository \footnote{\url{https://github.com/Zahin-10/tpm-reset}}.

\subsection{Experimental Setup}

\paragraph{Target Systems} We selected two hardware configurations with discrete TPMs (Infineon SLB9670) to validate attack generalizability: (1) \textbf{Target Laptop}: Lenovo ThinkPad T480 with the TPM onboard, and (2) \textbf{Target Workstation}: ASRock B850 Pro motherboard with a TPM module connected to the SPI pin header. Both systems were configured with Windows 11 and BitLocker using TPM-only protectors bound to PCR 7 and PCR 11, Windows' default settings.

\paragraph{Attacker Capabilities and Setup} The attack was conducted using a Fedora-based live Linux environment booted from USB media. Two critical Microsoft Secure Boot certificates were downloaded from official Microsoft links: the Windows Production CA 2011 (used by UEFI firmware to validate first-party Windows bootloaders) and the Microsoft UEFI CA 2011 (used to validate third-party bootloaders including Linux distributions signed through Microsoft's third-party signing service).The Windows Production CA certificate is essential for reconstructing the PCR7 digest that BitLocker expects from a legitimate Windows boot sequence.

\paragraph{Gather "Almost Good" Event Logs} Before executing the attack, we extracted the TCG Event Log from the kernel's binary measurement log interface while booted into the live Linux environment. These logs are "almost good" because they accurately capture all firmware-level measurements up to the \texttt{EV\_SEPARATOR} event, but diverge at the critical \texttt{EV\_EFI\_VARIABLE\_AUTHORITY} event. This divergence occurs because the UEFI firmware uses different certificates to validate different boot paths: Microsoft's first-party Windows Production CA 2011 for Windows bootloaders versus Microsoft's third-party UEFI CA 2011 for Linux bootloaders signed through Microsoft's third-party signing service. Since BitLocker's PCR7 policy expects the Windows-specific certificate measurement, we cannot directly replay the Linux boot's authority event. To address this, we simulated the correct \texttt{EV\_EFI\_VARIABLE\_AUTHORITY} event during replay.

\subsection{Attack Implementation}

\paragraph{Hardware-Level TPM Reset} The core of our attack exploits the ability to reset discrete TPMs independently of the host platform. To prevent OS interference during reset, we issue a systematic driver unbind/rebind procedure. The TPM device can be identified through kernel message logs and is unbound from its driver via the platform bus interface. Accessing the necessary pin on on the Target Laptop was not straightforward because the TPM is located on the keyboard-facing side of the mainboard. However, using schematics of the laptop found online, we identified the M.2 connector's \texttt{PERST\#} pin to connect to the platform reset rail \texttt{PLTRST\_NEAR} that also controls the onboard dTPM reset lines. A brief ground connection to this pin while running the live OS causes TPM reinitialization without system reboot. On the Target Workstation, we interrupt the TPM module's VCC supply by disconnecting and reconnecting the power line. Both methods successfully force the TPM into the \texttt{Startup(CLEAR)} state, zeroing all PCR values. After performing the hardware reset, the driver is rebound to restore communication. Post-reset, the TPM appears as a new device node, confirming successful reinitialization.

\paragraph{Replay Gathered Events with Deviating PCR7} The replay attack systematically reconstructs the measurement chain by replaying events from the TCG Event Log that can be extracted from Linux up to the \texttt{EV\_SEPARATOR} event, then manually constructing the correct Windows-path \texttt{EV\_EFI\_VARIABLE\_AUTHORITY} event. We programmatically build the \texttt{UEFI\_VARIABLE\_DATA} structure by concatenating five precise fields: the target system's CurrentPolicy GUID (16 bytes), UnicodeNameLength (8 bytes), VariableDataLength (8 bytes), the UTF-16 encoded variable name \texttt{L"db"}, and the \texttt{EFI\_SIGNATURE\_DATA} payload corresponding to Microsoft's Windows Production CA 2011 certificate. This complete binary structure is then hashed with SHA-256 to produce the exact digest that would be measured during a legitimate Windows boot. By extending PCR7 with this reconstructed digest rather than the Linux-boot digest from our captured log, we successfully forge the PCR7 value that BitLocker expected, enabling the subsequent unsealing operation.

\paragraph{Unseal Disk Key} Once PCR values are forged to match the reference boot state, we extract the BitLocker sealed object metadata from the volume header using \texttt{dislocker-metadata} to obtain the TPM2B\_PUBLIC and TPM2B\_PRIVATE structures required for unsealing. We then initiate the unsealing process by creating a TPM policy session, followed by applying \texttt{PolicyAuthValue} and \texttt{PolicyPCR} validation against the reconstructed PCR state. The \texttt{TPM2\_Unseal} operation successfully releases the sealed data, from which the 32-byte BitLocker Volume Master Key is extracted.

\paragraph{Mount Encrypted Drive and Decrypt} The extracted Volume Master Key is used to unlock and mount the BitLocker volume using \texttt{dislocker-fuse}. This process provides complete read-write access to the previously encrypted Windows volume.

\subsection{Results and Validation}

\paragraph{Results} The attack successfully compromised both target systems, demonstrating that the event log replay with reconstructed \texttt{EV\_EFI\_VARIABLE\_AUTHORITY} reliably forges the required PCR state for BitLocker unsealing. The deterministic nature of the attack ensures reproducibility: given the correct CurrentPolicy GUID and Microsoft certificates, the reconstructed PCR7 digest will always match the expected value from a legitimate Windows boot. Complete volume decryption was achieved in under 10 minutes of physical access time on both platforms, with the majority of time spent on hardware manipulation and driver rebinding rather than the computational replay process.

\paragraph{Validation} After mounting with \texttt{dislocker-fuse}, we verified complete read-write access to the Windows system files, user documents, and application data, confirming that the unsealed Volume Master Key correctly decrypted the BitLocker volume. The ability to browse, read, and modify files on the previously encrypted volume demonstrated successful circumvention of the TPM-based protection.

\paragraph{Limitations and Scope} The attack assumes a BitLocker configuration with a TPM-only protector. Systems using TPM+PIN or additional authentication factors remain protected against this vector. While our experiments focused on BitLocker, this attack scenario is equally applicable to LUKS-encrypted systems using TPM2 unsealing; however, LUKS lacks default PCR configurations and is entirely user-configurable, requiring attack-specific PCR reconstruction based on the target's custom binding policy. Platform dependency varies; some designs may couple TPM reset to system reset, preventing independent TPM manipulation. However, our successful validation on two distinct architectures suggests broad applicability across discrete TPM integrations.

\section{Discussion}

\subsection{Bus Protection Techniques}
\label{sec:bus_protection}

\paragraph{In-kernel TPM Session Hardening} Linux implements additional defense-in-depth measures by mandating HMAC sessions for all kernel TPM commands and using a \emph{null primary} derived from the TPM's null seed as session salt~\cite{linux-kernel-tpm-security}. Since the null seed changes across TPM resets, sessions salted under a pre-reset context fail post-reset, potentially detecting reset/replay attempts through "trust handoff" mechanisms between boot components. In our attack, however, the Linux kernel used is entirely attacker-controlled, so this feature can be disabled for the course of the attack.

\paragraph{Parameter Encryption and Mutual Authentication} TPM 2.0 provides confidentiality through \emph{parameter encryption} within authorization sessions rather than link-layer protection~\cite{tcg-tpm2-part1}. When a session specifies a symmetric algorithm, the TPM and caller derive a session key using Key Derivation Functions that mix nonces and shared secrets~\cite{tcg-tpm2-part1}. Command parameters marked \emph{decrypt} are encrypted under this session key, protecting sensitive payloads from passive bus observation while leaving command headers and handles visible~\cite{tcg-tpm2-part3}. HMAC-based mutual authentication provides integrity protection for protected parameters, detecting tampering during transit~\cite{tcg-tpm2-part1}.

\paragraph{Effectiveness Against Reset/Replay} While parameter encryption and mutual authentication effectively counter passive sniffing attacks, they provide \textbf{no protection against the reset/replay technique demonstrated in this work}. Our attack operates entirely within an attacker-controlled environment where legitimate TPM sessions can be established after hardware reset. The fundamental vulnerability lies not in bus communication security, but in the deterministic nature of measurement chains combined with the hardware reset capability -- a gap that current TCG bus protection guidance does not address~\cite{tcg_cpu_tpm_bus_protection_active_2023}.

\paragraph{Firmware TPM Integration} The most effective architectural defense against bus attacks is eliminating external communication channels entirely through firmware TPM (fTPM) implementations~\cite{tcg_cpu_tpm_bus_protection_passive_2023,tcg_cpu_tpm_bus_protection_active_2023}. By integrating TPM functionality within the CPU or chipset, fTPMs remove externally accessible traces that enable both passive sniffing and active manipulation. This architectural approach complements session-level protections and significantly reduces the attack surface compared to discrete TPM implementations~\cite{raj_ftpm_2016}.

\subsection{BIOS Password Protection}
\label{sec:bios_protection}

BIOS password protection might seem like a logical first-line defense against booting into alternative boot media like the attacker-controlled Linux. However, empirical analysis reveals that BIOS passwords constitute a shallow mitigation that fails to meaningfully increase attack cost or complexity. BIOS authentication operates during the Boot Device Selection (BDS) phase, creating fundamental attack windows through hardware bypasses and pre-authentication firmware vulnerabilities~\cite{uefi-spec-2025}. The most direct bypass leverages CMOS configuration storage, where BIOS passwords are typically stored alongside system configuration. Simple CMOS clear jumper manipulation requires no tools beyond basic jumper cables~\cite{dell-bios-clear-2024}, while I2C bus interruption techniques can reset passwords on consumer laptops using readily available hardware~\cite{cybercx-bios-bypass-2024}. These techniques exploit the physical accessibility of password storage mechanisms that must remain available for legitimate system operation. Firmware vulnerabilities in SEC, PEI, or DXE phases enable code execution before BIOS password authentication occurs. Examples include CVE-2025-3052 (NVRAM manipulation in DXE phase) and LogoFAIL vulnerabilities (BMP parsing flaws in graphics drivers)~\cite{binarly-nvram-vuln-2025,binarly-logofail-2023}. Since these vulnerabilities execute before password prompts, they completely bypass authentication mechanisms while maintaining the same economic profile as the reset/replay attack.

\subsection{PCR Secrecy as a Countermeasure}
\label{sec:pcr_secrecy}

The reset/replay attack leverages host-visible telemetry — specifically the TCG event log and direct \texttt{TPM2\_PCR\_Read} access — to reconstruct target PCR digests and groom the TPM to matching states prior to unsealing~\cite{edk2-trusted-boot,tcg-pfp-2023}. This raises the question: could hiding PCR values and event logs from the host OS through SMM mediation, privileged filtering, or architectural changes provide meaningful protection?

Limiting host access to PCR values would remove an oracle that reveals target digests and extend sequences, increasing the difficulty of deterministic PCR grooming~\cite{tcg-tpm2-part1}. However, PCR secrecy creates significant operational challenges. TCG attestation expects accompanying event logs to enable verifier explanation and failure diagnosis~\cite{tcg-pfp-2023}, while platform components routinely read PCRs for policy decisions, health monitoring, and update staging. Moreover, secrecy provides only friction rather than fundamental protection, as privileged adversaries who can influence measurements may still groom PCRs through captured traffic or reproduced extend sequences~\cite{hacky-solutions-tpm-attack-2024}.

\subsection{Sniffing vs. Reset/Replay}
\label{sec:attack_comparison}

The TCG claims that "passive attacks typically require less effort to mount than active attacks" \cite{tcg_cpu_tpm_bus_protection_passive_2023}.
However, we argue that our demonstrated reset attack is even easier to conduct than a sniffing attack:

\paragraph{Hardware Cost and Accessibility} Bus sniffing attacks require significant upfront investment: professional logic analyzers cost \$1,499~\cite{saleae-logic-pro-16}, budget FPGA solutions cost \$97~\cite{lattice-icestick-2024}, and specialized microcontroller approaches cost approximately \$10~\cite{stacksmashing-tpm-attack}. In contrast, the reset/replay attack demonstrated in this work requires only basic jumper cables (\$0-5), representing a 98.5-100\% cost reduction compared to sniffing approaches. This eliminates economic barriers entirely, making TPM circumvention accessible with minimal hardware investment.

\paragraph{Technical Prerequisites and Scalability} Both attacks demand board-level reverse engineering to locate TPM power lines on each target system.
However, sniffing requires precise probe attachment for all bus lines without permanent damage.
The reset/replay technique only requires finding either the TPM chip's supply voltage rail or a respective reset line.
Once a live OS is booted, it can be executed entirely in software using standard TPM tools and event log parsing, making it far more scalable for opportunistic attacks such as border searches or device theft scenarios.

\section{Conclusion}
With the ability to boot from an attacker-controlled system and brief physical access to the motherboard, our demonstrated attack against discrete TPMs 2.0 was able to break the full disk encryption of both a modern laptop and a workstation in minutes.
This shows that TPM-only protection schemes backed by dTPMs are fundamentally inadequate against physical adversaries. 
Whether through bus sniffing or reset-replay techniques, discrete TPM protection can be circumvented by attackers with minimal resources and technical expertise. This vulnerability stems from the architectural decision that prioritizes measurement transparency over replay resistance and therefore continues to exist with TPM 2.0.
To mitigate this, we suggest upgrading existing dTPM-based systems to use an additional factor like a PIN or password.
New system designs should integrate TPMs into the main CPU or chipset, thereby rendering active or passive attacks against the bus infeasible.

\section*{Acknowledgment}
This research was funded by the German Federal Ministry of Research, Technology and
Space within the DI-SIGN-HEP Project Grant 16KIS2074.

\bibliographystyle{IEEEtran}
\bibliography{references}

\end{document}